\documentclass[12pt]{iopart}

\usepackage{graphicx}
\usepackage{cite}

\eqnobysec

\begin{document}


\title[Grafted polymers on 3d SG fractals]
{Exact study of surface critical exponents of polymer chains grafted to adsorbing boundary of fractal lattices embedded in three-dimensional space}

\author{I \v Zivi\'c\dag, S Elezovi\'c-Had\v zi\'c\ddag~and S Milo\v sevi\'c\ddag}

\address{ \dag Faculty of Science, University of Kragujevac,  R. Domanovi\'ca 12, 34000~Kragujevac, Serbia}
\address{\ddag
University of Belgrade, Faculty of Physics, P.O.Box 44, 11001~Belgrade, Serbia}

\eads{\mailto{ivanz@kg.ac.rs}, \mailto{suki@ff.bg.ac.rs},
\mailto{savam@ff.bg.ac.rs}}

\begin{abstract}

We study the  adsorption problem of linear polymers, when the container of the polymer--solvent system is taken to be a member of the three dimensional Sierpinski gasket (SG) family of fractals. Members of the SG family are enumerated by an integer $b$ ($2\le b\le\infty$), and it is assumed that one side of each SG fractal is impenetrable adsorbing boundary. We calculate the critical exponents $\gamma_1, \gamma_{11}$, and $\gamma_s$ which, within the self--avoiding walk model (SAW) of polymer chain, are associated with the numbers of all possible SAWs with one, both, and no ends grafted  on the adsorbing impenetrable boundary, respectively. By applying the exact renormalization group (RG) method, for $2\le b\le 4$,  we have obtained specific values for these exponents, for various type of  polymer conformations. We discuss their mutual relations and  their relations with other critical exponents pertinent to SAWs on the SG fractals.
\end{abstract}


\vskip 5mm



\maketitle

\section{Introduction}
\label{prva}

The statistical properties of linear polymers near an impenetrable short--range attractive boundary have been extensively studied for a long time. The most frequently applied model for a polymer chain has been the self--avoiding walk (SAW) model (that is, the walk without self--intersections), so that steps of the walk have been identified with monomers that comprise the polymer, while the solvent surrounding has been represented by a lattice. Here we assume that polymer is immersed in a good solvent, and interacts only with an adsorbing surface bounding the polymer container, so that for strong enough  monomer-surface interaction the polymer undergoes phase transition from desorbed to adsorbed phase.

Since the polymer adsorption is a surface critical phenomenon, it has been possible to describe various polymer quantities in terms of power laws described by concomitant critical exponents.  Early investigations of polymer behavior  near attractive surfaces dealt with polymer chains immersed in homogeneous  spaces with planar adsorbing boundaries (see \cite{Eisenriegler} for a review).  These  studies  have been subsequently extended to polymers immersed in porous (inhomogeneous) media,  modeled by fractal lattices embedded in two-dimensional \cite{kumar1,z94,z95} and three-dimensional \cite{Bouchaud,z15} space. In these studies, almost exclusively, only two critical exponents have been studied, that is, the end--to--end distance critical exponent $\nu$ and the crossover exponent $\phi$ (that governs the number of contacts between the polymer and the surface). However, a complete  picture about the adsorption problem requires knowledge of surface critical exponents that describe numbers of polymer configurations grouped according to the different ways of anchoring to the adsorbing boundary. In terms of the self--avoiding random walk (SAW) model of linear polymers, these exponents are defined by the following formulas for numbers of possible different configurations averaged over the number of sites on the impenetrable surface
\begin{equation}\label{defgamma}
\fl
C_{11}(N,T)\sim\mu^NN^{\gamma_{11}-1}\>,\quad C_1(N,T)\sim\mu^NN^{\gamma_1-1}\>,\quad
C_s(N,T)\sim\mu^NN^{\gamma_s-1}\>,
\end{equation}
valid for large number $N$ of SAW steps. Here   $C_{11}, C_1$, and $C_s$, are numbers of all possible SAWs  with  both, one, and no ends grafted on the boundary
respectively,  $\mu=\mu(T)$ is temperature dependent connectivity constant and
$\gamma_{11}, \gamma_{1}$, and $\gamma_s$, are  concomitant  surface critical exponents that take different values in various polymer phases. So far, surface critical exponents have been  studied  mostly for SAWs  near the boundary surfaces of  two and tree-dimensional Euclidean spaces. These studies  were performed   using various techniques including series enumeration  \cite{staro1, vanderzande,foster}, conformal invariance theory \cite{staro2,cardy}, Coulomb gas method \cite{Duplantier2}, field theoretical approach \cite{diehl94, diehl98}, and Monte Carlo simulations \cite{grassberger94, grassberger05}.
On fractals, the  surface critical exponents where studied only for SAWs
immersed in a good solvent on two-dimensional fractal lattices \cite{Bubanja,EKMZ}. In this paper we  study the surface  critical exponents  for the polymer chain  situated on fractals that belong to the three-dimensional (3d) Sierpinski gasket (SG) family. Each member of the SG family is labeled by an integer $b$ ($2\le b\le\infty$), and it is assumed that one side of each SG fractal is impenetrable adsorbing wall. By applying an exact renormalization group (RG) method for the SAW model that includes monomer-surface interactions we have calculated critical exponents $\gamma_{11}$, $\gamma_1$, and $\gamma_s$, for $b=2$, 3 and 4 fractals.

This paper is organized as follows. In section~\ref{druga} we describe the 3d SG fractals for general scaling parameter $b$, and introduce the self-avoiding walk model in the case when a  boundary of 3d SG fractal is an adsorbing surface. Then, we
present the framework of the general RG method for studying the polymer adsorption problem on these fractals. In section~\ref{rezultati} we display  the exact results for the studied critical exponents $\gamma_1, \gamma_{11}$, and $\gamma_s$ for $b=2$, 3 and 4 fractals, in different polymer regimes. All obtained results are summarized, discussed and  compared with related previous results in  section~\ref{diskusija}. Finally, some technical details are given in the Appendix.

\section{Framework of the renormalization group approach}\label{druga}

In this section we are going to expound on the renormalization group~(RG) approach of calculating the critical exponents $\gamma_1, \gamma_{11}$, and $\gamma_s$ for the  adsorption problem of SAWs immersed in a solvent modeled by fractals belonging to the 3d SG family of fractals. Here we give a brief summary of their basic properties. We start with recalling the fact that each member of  3d SG fractal family is labeled by an integer $b\ge 2$ and can be constructed in stages. At the first stage ($r=1$) of the construction there is a tetrahedron of base $b$ containing $b(b+1)(b+2)/6$ upward oriented unit tetrahedrons. The subsequent fractal stages are constructed recursively, so that the complete self-similar fractal lattice can be obtained as the result of an infinite iterative process of successive $(r\to r+1)$ enlarging the fractal structure $b$ times, and replacing the smallest parts of enlarged structure with the initial ($r=1$) structure.
 In the case under study, we take that one
of the four boundaries of the 3d SG fractal is impenetrable adsorbing surface (wall), which is itself a 2d SG fractal with the fractal dimension
$d_s=\ln[b(b+1)/2]/\ln b$, whereas the fractal dimension of
the complete 3d SG fractal is  $d_f={{\ln [{{b(b+1)(b+2)}/6}}]/{\ln b}}$.

\begin{figure}
\hskip5cm
\includegraphics[scale=0.3]{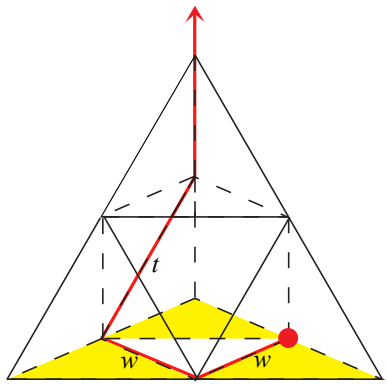}
\caption{The fractal structure of the $b=2$ 3d SG fractal at the
first stage of construction, with an example of the SAW path starting from the
adsorbing surface depicted by the yellow area. The steps on the adsorbing surface and in the
adjacent layer are weighted by the factors $w=e^{-\epsilon_w/k_BT}$
and $t=e^{-\epsilon_t/k_BT}$, respectively. Here $\epsilon_w$ is the
energy of a monomer lying on the adsorbing wall ($\epsilon_w<0$),
and $\epsilon_t>0$ is the energy  of a monomer that appears in the
layer adjacent to the wall. The depicted SAW path corresponds to a polymer configuration with one end (red point)  grafted on the adsorbing surface
and has the weight $x^4w^2t$.}
 \label{fig1}
\end{figure}

In order to describe the effect of attractive (adsorbing) surface, one
should introduce two Boltzmann factors: $w={\mathrm{e}}^{-\varepsilon_w/k_BT}$, and
$t={\mathrm{e}}^{-\varepsilon_t/k_BT}$, where $\varepsilon_w$ is the energy of a monomer
lying on the adsorbing surface, and $\varepsilon_t$ is the energy
of a monomer in the layer adjacent to the surface.
If we assign the weight $x$ to a single step of the SAW walker, then
the weight of a walk having $N$ steps, with $M$ steps on the surface, and $K$ steps in the layer
adjacent to the surface, is $x^Nw^Mt^K$  (see figure~\ref{fig1}).

The weighting factors defined in the foregoing paragraph
allow us to introduce the following global generating functions
\begin{eqnarray}
\fl G_{11}(x,T)&=&\sum_{N=1}^\infty{x^N}\sum_{M,K} {{\mathcal G}_{11}(N,M,K)
w^Mt^K}=\sum_{N=1}^\infty {C_{11}(N,T) x^N}\>\label{eq:G11}\>,\\
\fl G_{1}(x,T)&=&\sum_{N=1}^\infty{x^N}\sum_{M,K} {{\mathcal G}_{1}(N,M,K)
w^Mt^K}=\sum_{N=1}^\infty {C_{1}(N,T) x^N}\>\label{eq:G1}\>,\\
\fl G_{s}(x,T)&=&\sum_{N=1}^\infty{x^N}\sum_{M,K} {{\mathcal G}_{s}(N,M,K)
w^Mt^K}=\sum_{N=1}^\infty {C_{s}(N,T) x^N}\>\label{eq:Gs}\>,
\end{eqnarray}
where ${\mathcal G}_{1}(N,M,K)$ (${\mathcal G}_{11}(N,M,K)$) represents the average number (over all sites of adsorbing wall) of
$N$-step SAWs with  $M$ steps on the surface
and $K$ steps in the layer adjacent to the wall provided one (both)
 end(s) of the walk is (are) anchored to the wall,
while ${\mathcal G}_s(N,M,K)$ is the number of SAWs with no ends anchored to
the wall. If we assume that, for large $N$, the numbers $C_{11}(N,T)$, $C_{1}(N,T)$ and $C_{s}(N,T)$ behave in accordance with the laws defined in (\ref{defgamma}), then the leading singular behavior of the generating functions are of the form
\begin{equation}
\fl G_{11}(x,T) \sim (1-x\mu )^{-\gamma_{11}},\quad
G_{1}(x,T) \sim (1-x\mu )^{-\gamma_{1}},\quad
G_{s}(x,T) \sim (1-x\mu )^{-\gamma_{s}},  \label{eq:powerlaw}
\end{equation}
in the vicinity of the critical value $x_c=1/\mu$, {\it i.e.}  when $x$ approaches $x_c$ from below.

\begin{figure}[b]
\hskip2cm
\includegraphics[scale=0.4]{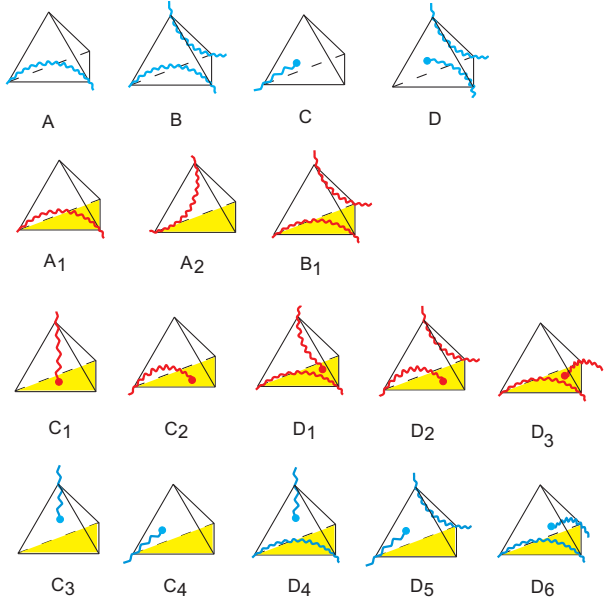}
\caption{Schematic representation of traversing and one-leg restricted  generating functions used to describe all possible   configurations of SAW (needed to construct  the global generating functions $G_{11}$, $G_{1}$ and $G_{s}$), within the $r$th stage of 3d SG fractal structure. Thus, for example, the $C_1^{(r)}$ represents the situation when SAW starts from the impenetrable surface  and leaves the $r$th stage fractal lattice at its vertex which does not lie in the surface.  The interior details of the $r$th stage fractal structure, as well as  details of the chains, are not shown. Small red circles depict the SAW starting points from the attaching surface, while the blue ones denote the SAW starting points that are in the bulk part  of tetrahedron lying on the  surface.} \label{fig2} \end{figure}

To calculate the surface critical exponents $\gamma_1, \gamma_{11}$, and $\gamma_s$, we have found that it is helpful to define three kinds of  restricted generating functions that provide a complete description of the generating functions $G_{11}$, $G_{1}$ and $G_{s}$. These function  are: the traversing SAW generating functions $A^{(r)}$,  $B^{(r)}$,
 $A_1^{(r)}$, $A_2^{(r)}$ and $B_1^{(r)}$;
one-leg SAW functions $C^{(r)}$, $D^{(r)}$, $C_i^{(r)}\, (i=1,\ldots,4)$ and $D_i^{(r)}\, (i=1,\ldots,6)$ (which are depicted in figure~\ref{fig2}); and, the two-leg SAW functions
$E_m^{(r)}$ (which are not presented in figure~\ref{fig2}, because they are not relevant for the critical exponent calculation~\cite{Dhar78}). Using these restricted generating functions, one can express each global generating function in the following manner. We start with the function $G_{11}$, that can be written in the form
\begin{eqnarray}
\fl G_{11}(x,T)=\sum_{r=0}^{\infty}{1\over
{\left[{b(b+1)\over 2}\right]}^{r+1}}
&\Biggl(&\sum_{i=1}^{2}\sum_{j=1}^{2}\>f_{ij}C_i^{(r)}C_j^{(r)}+
\sum_{i=1}^{2}\sum_{j=1}^{3}\>g_{ij}C_i^{(r)}D_j^{(r)}
\nonumber\\
&+&\sum_{i=1}^{3}\sum_{j=1}^{3}\>h_{ij}D_i^{(r)}D_j^{(r)}+
\sum_{m}e_m E_m^{(r)}\Biggr)\>,
 \label{g11}
 \end{eqnarray}
where  the coefficients $f_{ij}$, $g_{ij}$, $h_{ij}$ and $e_m$  are
polynomials in $A^{(r)},B^{(r)},A_1^{(r)},A_2^{(r)}$ and $B_1^{(r)}$.
Similarly, the generating functions $G_1(x,T)$ and $G_s(x,T)$
can be written as
\begin{eqnarray}
\fl G_{1}(x,T)&=&G_{11}(x,T)+\sum_{r=0}^{\infty}{1\over
{\left[{b(b+1)\over 2}\right]}^{r+1}}
\Biggl[\sum_{i=1}^{2}\sum_{j=3}^{4}\>p_{ij}C_i^{(r)}C_j^{(r)}+
\sum_{i=1}^{2}\sum_{j=4}^{6}\>q_{ij}C_i^{(r)}D_j^{(r)}
\nonumber\\
\fl &+& \sum_{i=3}^{4}\sum_{j=1}^{3}\>r_{ij}C_i^{(r)}D_j^{(r)}
+ \sum_{i=1}^{3}\sum_{j=4}^{6}\>s_{ij}D_i^{(r)}D_j^{(r)}+
\sum_{i=1}^{2}C_i^{(r)}(p_{i}C^{(r)}+q_{i}D^{(r)})\nonumber\\
\fl&+&
\sum_{i=1}^{3}D_i^{(r)}(r_{i}C^{(r)}+s_{i}D^{(r)})
+\sum_{m}t_m E_m^{(r)}\Biggr]\>,
 \label{g1}
 \end{eqnarray}
and
\begin{eqnarray}
\fl G_{s}(x,T)&=&\sum_{r=0}^{\infty}{1\over
{\left[{b(b+1)\over 2}\right]}^{r+1}}
\Biggl[\sum_{i=3}^{4}\sum_{j=3}^{4}\>p'_{ij}C_i^{(r)}C_j^{(r)}+
\sum_{i=3}^{4}\sum_{j=4}^{6}\>q'_{ij}C_i^{(r)}D_j^{(r)}
\nonumber\\
\fl &+&
 \sum_{i=4}^{6}\sum_{j=4}^{6}\>s'_{ij}D_i^{(r)}D_j^{(r)}+
\sum_{i=3}^{4}C_i^{(r)}(p'_{i}C^{(r)}+q'_{i}D^{(r)})
+\sum_{i=4}^{6}D_i^{(r)}(r'_{i}C^{(r)}+s'_{i}D^{(r)})
\nonumber\\
\fl&+&p'(C^{(r)})^2+q'C^{(r)}D^{(r)}+s'(D^{(r)})^2+
\sum_{m}t'_m E_m^{(r)}\Biggr]\>,
 \label{gs}
 \end{eqnarray}
where, again, the
coefficients  standing with one-- and two--leg generating functions on the right-hand side, are polynomials in traversing functions $A^{(r)},B^{(r)},A_1^{(r)},A_2^{(r)}$ and $B_1^{(r)}$.

Due to the self-similarity of fractals, restricted generating functions obey recursive relations, which can be interpreted as RG equations~\cite{Dhar78}.  For arbitrary $r$,  these equations for
the bulk restricted partition functions
(for any $b\geq 2$) have the form
\begin{eqnarray}
\fl A^{(r+1)}=\sum_{i,j}a(i,j) \left(A^{(r)}\right)^i\left(B^{(r)}\right)^{j}  \,,
\label{eq:RGA}
\\
\fl B^{(r+1)}=\sum_{i,j}b(i,j) \left(A^{(r)}\right)^i\left(B^{(r)}\right)^{j} \,,
\label{eq:RGB}\\
\fl C^{(r+1)}=c_1(A^{(r)},B^{(r)})C^{(r)}+c_2(A^{(r)},B^{(r)})D^{(r)},\label{eq:RGC} \\
\fl D^{(r+1)}=d_1(A^{(r)},B^{(r)})C^{(r)}+d_2(A^{(r)},B^{(r)})D^{(r)},\label{eq:RGD}
\end{eqnarray}
where $a(i,j)$ and $b(i,j)$ are non-negative integers, whereas coefficients
$c_1(A,B)$, $c_2(A,B)$, $d_1(A,B)$ and  $d_2(A,B)$ are polynomials in $A$ and $B$, neither of them depending on $r$~\cite{DharVannimenus,Knezevic87,z15}. For surface traversing functions, $A_1^{(r)}, A_2^{(r)}$ and $B_1^{(r)}$,  RG equations have the form
\begin{eqnarray}
\fl A_1'=\sum_{i,j,i_1,i_2,j_1}
 a_1(i,j,i_1,i_2,j_1)\, A^{i}\,B^{j}\,
A_1^{i_1}\,A_2^{i_2}\,B_1^{j_1}  ,  \label{eq:RGA1} \\
\fl A_2'=\sum_{i,j,i_1,i_2,j_1}
 a_2(i,j,i_1,i_2,j_1)\, A^{i}\,B^{j}\,
A_1^{i_1}\,A_2^{i_2}\,B_1^{j_1} , \label{eq:RGA2} \\
\fl B_1'=\sum_{i,j,i_1,i_2,j_1}
 b_1(i,j,i_1,i_2,j_1)\, A^{i}\,B^{j}\,
A_1^{i_1}\,A_2^{i_2}\,B_1^{j_1} , \label{eq:RGB1}
\end{eqnarray}
where on the left-hand side we have used the prime symbol as a superscript for the $(r+1)$-th order generating functions, and no indices on the right-hand side for the $r$-th order functions. The numbers $a_1$, $a_2$, and $b_1$ do not depend on $r$~\cite{Bouchaud,z15}. Analogously, we can construct  additional recursion relations for surface one-leg generating functions
\begin{equation}\label{c1d6}
 \left[ \matrix{C_1'\cr C_2'\cr D_{1}'\cr D_{2}'\cr
D_{3}'}\right]= {\mathcal M_S}\left[ \matrix{C_1\cr C_2\cr D_{1}\cr D_{2}\cr
D_{3}}\right]\, , \quad \qquad
\left[ \matrix{C_3'\cr C_4'\cr D_{4}'\cr D_{5}'\cr
D_{6}'}\right]= {\mathcal M_S}\left[ \matrix{C_3\cr C_4\cr D_{4}\cr D_{5}\cr
D_{6}}\right]+{\mathcal M_B}\left[ \matrix{C\cr D\cr 0\cr 0 \cr
0}\right]\>,
\end{equation}
where elements of matrices ${\mathcal M_S}$ and ${\mathcal M_B}$ ($S$ denotes the surface functions, while $B$ the bulk ones) are polynomials  in  $A$, $B$, $A_1$, $A_2$ and $B_1$ functions. Starting with the initial conditions
\begin{equation}
A^{(0)}=x\, , \quad A_1^{(0)}=wx\, , \quad A_2^{(0)}=tx\, ,\quad
 B^{(0)}=x^2 \, , \quad
B_1^{(0)}=wtx^2\, ,\label{eq:PocetniUslovi}
\end{equation}
which correspond to the elementary tetrahedron $(r=0)$\footnote{Here we note that in the approach applied in the present paper the SAW is forced to leave the unit tetrahedron after completing one step. Such restriction simplifies the model, but is not expected to alter the critical behavior~\cite{DharVannimenus}.} one
can iterate RG relations
(\ref{eq:RGA}),  (\ref{eq:RGB}), and    (\ref{eq:RGA1})-(\ref{eq:RGB1}) for traversing SAW function in order to establish the phase diagram of the polymer system. In addition, we define the initial condition for one-leg generating functions
\begin{eqnarray}
 C^{(0)}=1+3x, \quad D^{(0)}=x+x^2, \nonumber\\
 C^{(0)}_1=3xt, \quad C^{(0)}_2=1+2xw, \quad C^{(0)}_3=1, \quad C^{(0)}_4=xt\>,\nonumber\\
 D^{(0)}_1=x^2wt, \quad D^{(0)}_2=xt+x^2wt, \quad D^{(0)}_3=xw, \nonumber\\
 D^{(0)}_4=xw\, , \quad D^{(0)}_5=0 , \quad D^{(0)}_6=x^2wt,
\label{eq:PocetniUsloviLeg}
\end{eqnarray}
that are needed for  studying large $r$ behavior of these functions (through relations (\ref{eq:RGC}), (\ref{eq:RGD}) and (\ref{c1d6})), and consequently to determine the surface critical exponents $\gamma_{11}$, $\gamma_1$ and $\gamma_s$, from the singular parts of   functions $G_{11}$, $G_{1}$ and $G_{s}$. To perform described procedure we need to know all  RG equations for a specific fractal.
We have been able to complete  the exact form of required RG transformations and carry out a comprehensive analysis for  the first three members ($b=2$, 3 and 4)  of the 3d SG family of fractals.  The RG equations (\ref{eq:RGA})-(\ref{eq:RGB1}) were found in previous studies \cite{Bouchaud,z15,Dhar78,DharVannimenus,Knezevic87,z17}, and in this work we have determined the additional  RG transformation (\ref{c1d6}) by applying an exact enumeration method. For $b=2$ fractal these are given in the Appendix, while for $b=3$ and 4 fractals they can be obtained upon request to the authors (to be more precise, for the critical exponents calculation only  transformations defined with ${\mathcal M_{S}}$ were needed, and consequently only  that transformations are quoted). We note that computer enumeration and classification of all SAW configurations required to build RG transformations for one-leg generating functions, was done in a few  seconds for $b=2$ and $b=3$ fractals, while in $b=4$ case it took 7 hours on a PC with i5 Intel microprocessor. Details of the performed RG analysis together with the  specific results for $b=2$, 3 and 4 fractals   are presented  in the next section.

\section{Results for $b=2$, $3$ and $4$ fractals}\label{rezultati}

Numerical analysis  of RG  equations for traversing SAW functions  showed that for each value of $t$ (between 0 and 1) there exists a critical value of $w = w_c(t)$, such that for values of $w$ smaller than $w_c(t)$ polymer is in desorbed state, whereas for $w>w_c(t)$ it is adsorbed at the surface.  Precisely at the critical value $w=w_c(t)$ the transition from adsorbed to desorbed phase occurs. In the following subsections all established polymer regimes will be reviewed separately and for each of them  the surface critical exponents will be evaluated.

\subsection{Desorbed phase $(w<w_c(t))$}

 For weak monomer-surface interactions $w<w_c(t)$, and critical value of the fugacity $x=x_c$ (which does not depend on the values of $w$ and $t$) the parameters ($A^{(r)}, B^{(r)}, A_1^{(r)}, A_2^{(r)}, B_1^{(r)}$)
tend to $(A^*, B^*,0,0,0)$, when $r\to\infty$, which indicates that
polymer, stays away from the attractive surface~\cite{Bouchaud,z15}. This state is referred to as desorbed
phase, determined by the RG fixed point
\begin{equation}
(A,B,A_1,A_2,B_1)^*=(A^*,B^*,0,0,0)\> .
\label{eq:DesorbedSAWFP}
\end{equation}
The mean squared end-to-end distance $\langle R_N^2\rangle$ of the polymer chain scales with its length $N$ as $N^{2\nu}$, where the  critical exponent $\nu$ is equal to
\begin{equation}
\nu=\frac{\ln b}{\ln \lambda_\nu}\, ,\label{eq:formulani}
\end{equation}
and $\lambda_\nu$ is the largest eigenvalue of the RG transformation (\ref{eq:RGA}) and (\ref{eq:RGB}) linearized in the vicinity of the corresponding fixed point $(A^*,B^*)$~\cite{Dhar78,Knezevic87,z15}.

In order to calculate  surface critical exponents $\gamma_{11}$, $\gamma_1$ and $\gamma_s$,
one should investigate singular behavior of the
generating functions (\ref{eq:G11})--(\ref{eq:Gs}), for which it is first necessary to analyze RG transformations (\ref{eq:RGA1})--(\ref{c1d6}) in
the vicinity of the bulk fixed point (\ref{eq:DesorbedSAWFP}).
After large number of RG iterations,
 the traversing functions  behave as
\begin{eqnarray}
\fl && A^{(r)}\approx A^*, \quad B^{(r)}\approx B^*, \quad
A_1^{(r)}\approx 0\,, \quad A_2^{(r)}\sim q^r\to 0,\quad  B_1^{(r)}\approx  0\,\label{d1},
\end{eqnarray}
where $q=\left({\partial A'_2\over \partial A_2}\right)^{*}$  is  irrelevant eigenvalue of (\ref{eq:RGA1})--(\ref{eq:RGB1}) calculated  at the fixed point (\ref{eq:DesorbedSAWFP}). Also, one-leg generating functions  have the following large $r$ behavior
\begin{eqnarray}
\fl && C^{(r)}\sim D^{(r)}\sim \lambda_B^r\,,\label{d20}\\
\fl &&C_1^{(r)}\sim \lambda_S^r, \quad C_2^{(r)}\sim  \mathrm{const}+(q\lambda_S)^r,\quad D_1^{(r)}\sim D_2^{(r)}\sim (q^2\lambda_S)^r,  \quad  D^{(r)}_3\sim (q^3\lambda_S)^r,
\label{d21}\\
\fl  &&  C_3^{(r)}\sim \lambda_B^r,  \quad  C_4^{(r)}\sim \mathrm{const}+(q\lambda_B)^r, \quad  D_4^{(r)}\sim D_5^{(r)}\sim (q^2\lambda_B)^r,  \quad  D^{(r)}_6\sim (q^3\lambda_B)^r,\label{d2}\end{eqnarray}
where $\lambda_B$ is the relevant eigenvalue of the matrix
\begin{equation}\label{lambdab}
{\mathcal M_{\gamma}}=\left[\matrix{c_1(A^*,B^*)&c_2(A^*,B^*)\cr
	d_1(A^*,B^*)&d_2(A^*,B^*)\cr}\right]  \>,
\end{equation}
 made out of the coefficients $c_i$ and $d_i$  appearing in RG equations (\ref{eq:RGC}) and (\ref{eq:RGD}) for the pure bulk functions $C^{(r)}$ and $D^{(r)}$, and
\begin{equation}\label{lambdas}
\lambda_{S}=
\left[{\mathcal M_{S}}\right]_{11}^*\>,
\end{equation}
asterisk denoting that the value of the polynomial $\left[{\mathcal M_{S}}\right]_{11}$ (appearing in RG equations (\ref{c1d6})) is evaluated in the fixed point (\ref{eq:DesorbedSAWFP}). We remind here that the eigenvalue $\lambda_B$ determines the bulk critical exponent
\begin{equation}\label{gamabulk}
\gamma={\nu\over\ln b}\,{{\ln{{\lambda_B^2}\over b(b+1)(b+2)/6}}}\>,
\end{equation}
which governs the singular behavior $G_{sing} \sim (1-x\mu )^{-\gamma}$
of the generating function
\begin{equation}\label{G0}
G(x,T)=
\sum_{N=1}^\infty {C(N,T)\, x^N}
\end{equation}
for all possible  SAWs in the bulk (away from adsorbing boundary),  where $C(N,T)\sim\mu^NN^{{\gamma}-1}$ is the average number (over all starting points) of such $N$-step SAWs~\cite{z17}.

From (\ref{d21}) and (\ref{d2}), we  perceive that, for  large $r$,  behavior of  $C_2^{(r)}$,  $C_4^{(r)}$ and $D_i^{(r)}~(i=\overline{1,6})$,  depends on mutual relation between the specific values of $\lambda_S$, $\lambda_B$ and $q$.  Calculated values of  $\lambda_S$ and $\lambda_B$, for $b=2$, 3 and 4 fractals,  are given in table~\ref{tabela1}, while the particular values for $q$  are: $q{(b=2)}=0.4294$, $q{(b=3)}=0.2285$ and $q{(b=4)}=0.1440$. Since for each studied fractal $\lambda_B>\lambda_S>1$ and $q^2<1/\lambda_B<q<1/\lambda_S$ one finds that $C_2^{(r)}\to$const, $C_3^{(r)}\gg C_1^{(r)}\to\infty$, $C_3^{(r)}\gg C_4^{(r)}\to\infty$, and $D_i^{(r)}\to 0$ $(i=\overline{1,6})$. Then, from (\ref{gs}) it follows that on the large scale $L=b^r$ behavior of the generating function $G_s^L$ is determined by the term containing $[C_3^{(r)}]^2$, and consequently
\begin{equation}
G_s^L\sim \left( \frac{\lambda_B^2}{\frac{b(b+1)}2}\right)^r\, .
\end{equation}
Since $L\sim \langle N\rangle^{\nu}$, and  $\langle N\rangle\sim (x_c-x)^{-1}$,
the critical behavior $G_s\sim (1-\mu x)^{-\gamma_s}$ follows, with
\begin{equation} \label{ekspBulkGs}
\gamma_{s}={\nu\over\ln b}\,{{\ln{{\lambda_B^2}\over b(b+1)/2}}}\, .
\end{equation}

By examining the large $r$ behavior of the terms in the sum (\ref{g11}), representing the generating function $G_{11}$, one concludes that the term containing $[C_1^{(r)}]^2$ dominates, so that the largest term in the sum behaves as $[\lambda_S^2/(b(b+1)/2)]^r$. However, $\lambda_S^2$ is smaller than $b(b+1)/2$ in each studied $b$ case, implying that $G_{11}$ remains finite at the critical point. Therefore, one should inspect the large scale behavior of the derivative ${\mathrm d  G_{11}}\over{\mathrm d x}$. By finding this derivative from (\ref{g11}), one finds that it can be expressed as an infinite sum, similar to the sum on right-hand side of (\ref{g11}), but with terms that, apart from RG parameters, also contain their derivatives with respect to $x$. On the other hand, by finding derivative of the RG equations  (\ref{eq:RGA})--(\ref{c1d6}), one can find recurrence relations for the RG parameters derivatives, and analyze their large $r$ behavior. From such an analysis, it follows that derivatives of the traversing RG parameters behave as $\lambda_\nu^r$, as well as ${\mathrm d  D_2^{(r)}}\over{\mathrm d x}$, whereas ${{\mathrm d  C_{1,2}^{(r)}}\over{\mathrm d x}}\sim (\lambda_S\lambda_\nu)^r$, ${{\mathrm d  D_{1}^{(r)}}\over{\mathrm d x}}\sim (q^2\lambda_S\lambda_\nu)^r$, and ${{\mathrm d  D_{3}^{(r)}}\over{\mathrm d x}}\to 0$. These findings, together with the established behavior of the RG parameters, imply that
\begin{equation}
\frac{\mathrm d  G_{11}^L}{\mathrm d x}\sim \left( \frac{\lambda_S^2\lambda_\nu}{\frac{b(b+1)}2}\right)^r\, ,
\end{equation}
with $\lambda_S^2\lambda_\nu> \frac{b(b+1)}2$ in all cases, meaning that the function $\frac{\mathrm d  G_{11}}{\mathrm d x}$ diverges on the large scale $L=b^r$. On the other hand, in the vicinity of the critical point, the following  relation~\cite{Bubanja} is satisfied
\begin{equation}
\frac{\mathrm d  G_{11}}{\mathrm d x}\sim (1-\mu x)^{-(\gamma_{11}+1)}\, ,
\end{equation}
whereupon follows
\begin{equation}\label{ekspBulkGama11}
\gamma_{11}={\nu\over\ln b}\,{{\ln{{\lambda_S^2}\over b(b+1)/2}}}\, .
\end{equation}

Finally, using the established behavior of the RG parameters and $G_{11}$,  from (\ref{g1}) one can, in a similar way, find that on large scale the generating function $G_1$ behaves as
\begin{equation}
G_1^L\sim \left( \frac{{\lambda_S}\lambda_B}{\frac{b(b+1)}2}\right)^r\, ,
\end{equation}
implying that
\begin{equation}
G_{1}\sim (1-\mu x)^{-\gamma_{1}}\, ,
\end{equation}
with
\begin{equation}\label{ekspBulkGama1}
\gamma_{1}={\nu\over\ln b}\,{{\ln{{\lambda_S\lambda_B}\over {b(b+1)/2}}}}\, .
\end{equation}

For all studied fractals, the obtained specific results of these critical exponents, in desorbed polymer phase, are listed in table~\ref{tabela1}.
{\Table{\label{tabela1}
Values of the critical exponents $\gamma_{11}$, $\gamma_1$ and $\gamma_s$ for desorbed $(w<w_c)$, attached $(w=w_c)$ and adsorbed $(w>w_c)$ phases, for $b=2$, 3, and 4 3d SG fractals. For the sake of completeness and comparison, here we also give coordinates of the corresponding fixed points (f.p.), the values of the end-to-end distance critical exponent $\nu$~\cite{z15,Dhar78,Knezevic87}  for desorbed phase $(w<w_c)$ and crossover region $(w=w_c)$, together with the corresponding eigenvalues  $\lambda_B$ and $\lambda_S$, as well as the  values of  $\gamma$ exponent  found previously in~\cite{z17}. For adsorbed SAW $(w>w_c)$ critical exponents $\gamma_{11}$, $\gamma_1$ and $\gamma_s$ are all equal to the value of $\gamma$ for the corresponding 2d SG fractal, calculated in~\cite{EKM}.}
 \begin{tabular}{llllllllll}
 \br
 \ms
\multicolumn{10}{c}{Desorbed  SAW   ($w<w_c$)  [ f.p. $(A^*,B^*,0,0,0)$ ] }\\
\ms
 $b$ & $A^*$&$B^*$&${\nu}$  & $\lambda_{B}$
 & $\lambda_S$& $\gamma$ &$\gamma_{11}$&$\gamma_{1}$&$\gamma_{s}$\\
\mr

2 &  0.4294&0.0500&0.6740&4.2069 &1.2883 &1.4461&-0.5756&0.5751&1.7259\\
3 &  0.3420&0.0239&0.6542&11.477&1.7206 &1.5352&-0.4206&0.7094&1.8394\\
4 &  0.2899&0.0122&0.6411&25.679 &2.3381 &1.6165&-0.2793&0.8289&1.9370\\
 \br
 \ms
 \multicolumn{10}{c}  {Surface attached  SAW $(w=w_c)$ [ f.p. $(A^*,B^*,A^*,A^*,B^*)$ ]}\\
 \ms
$b$ & $A^*$&$B^*$&${\nu}$  & $\lambda_{B}$
 & $\lambda_S$&  &$\gamma_{11}$&$\gamma_{1}$&$\gamma_{s}$\\
\mr
2 &  0.4294&0.0500&0.6740&4.2069 &3.1441 &&1.1596&1.4427&1.7259\\
3 &  0.3420&0.0239&0.6542&11.477&6.7759 &&1.2118&1.5256&1.8394\\
4 &  0.2899&0.0122&0.6411&25.679 &12.4404 &&1.2668&1.6019&1.9370\\
 \br
 \ms
 \multicolumn{10}{c}  {Adsorbed SAW   ($w>w_c$) [ f.p. $(0,0,A^{2d},0,0)$ ]}\\
 \ms
$b$ & $A^{2d}$&&${\nu}_{SAW}^{2d}$  & $\lambda_{B}^{2d}$
 &  &\multicolumn{4}{l}{$\gamma=\gamma_{11}=\gamma_{1}=\gamma_{s}$}\\
\mr
2 &0.6180  &&0.7986&3.1456&&\multicolumn{4}{l}{1.3752}\\
3 &0.5511  &&0.7936&6.6395&&\multicolumn{4}{l}{1.4407}\\
4 &0.5063  &&0.7884&11.645&&\multicolumn{4}{l}{1.4832}\\
 \mr
    \end{tabular}
\endTable}

\subsection{Adsorbed phase $(w>w_c(t))$} When $w$ is increased beyond
$w_c(t)$, and for $x=x_c(w,t)$~\cite{Bouchaud,z15} RG parameters flow towards the new fixed point that describes the adsorbed polymer chain
\begin{equation}
(A,B,A_1,A_2,B_1)^*=(0,0,A^{2d},0,0)\> ,
\label{eq:AdsorbedSAWFP}
\end{equation}
where $A^{2d}$ is the fixed point for the corresponding two--dimensional SG fractal~\cite{EKM}. In this fixed point polymer chain displays  the features of a polymer system situated on two-dimensional SG fractal, and consequently  all surface  critical exponents are equal to the corresponding   value of $\gamma$ for 2d SG fractals ({\it i.e.} $\gamma_1=\gamma_{11}=\gamma_s=\gamma(\hbox{2d SG})$), whose values were found in \cite{EKM}, and  are quoted  in table~\ref{tabela1}.

\subsection{Surface attached chain $(w=w_c(t))$}
Precisely at $w=w_c(t)$ behavior of the polymer abruptly changes. For this value of $w$, and $x=x_c$ (the same as for the desorbed phase), the
symmetric special fixed point is reached
\begin{equation}
 (A,B,A_1,A_2,B_1)^*=(A^*,B^*,A^*,A^*,B^*)\,,
\label{eq:symetric}
\end{equation}
which corresponds to the surface attached chain, when a balance between the attractive polymer--surface interaction and an
effective ``entropic" repulsion sets in. In this case, we have found the following  large $r$ behavior for  traversing functions
\begin{equation}
 A^{(r)}\approx A_1^{(r)}\approx A_2^{(r)}\approx A^*, \quad B^{(r)}\approx B_1^{(r)}\approx B^*, \label{funkcijeAB}\end{equation}
whereas the one-leg functions behave as
\begin{eqnarray}
 && C^{(r)}\sim D^{(r)}\sim \lambda_B^r\label{funkcijeCD0}\\
 &&C_1^{(r)}\sim C_2^{(r)}\sim D_1^{(r)}\sim D_2^{(r)}\sim D^{(r)}_3 \sim\lambda_S^r \,,
 \label{funkcijeCD1}\\
 &&C_3^{(r)}\sim C_4^{(r)}\sim D_4^{(r)}\sim D_5^{(r)}\sim D^{(r)}_6 \sim\lambda_B^r \,,\label{funkcijeCD3}\end{eqnarray}
where $\lambda_B$ has the same value as in desorbed SAW phase ({\it i.e.} it is the eigenvalue of (\ref{lambdab})),
and $\lambda_S$ is the largest eigenvalue of the matrix ${\mathcal M_{S}}$,  evaluated at the symmetric fixed point (\ref{eq:symetric}). Using already described approach, one can establish the same singular behavior of the generating functions with the relations (\ref{ekspBulkGs}),  (\ref{ekspBulkGama11}), and (\ref{ekspBulkGama1}), for the surface critical exponents  $\gamma_s, \gamma_{11}$, and $\gamma_1$, respectively, as in the case of the desorbed chain. The obtained specific results for  surface critical exponents, together with related  eigenvalues needed for their evaluation,  are given in table~\ref{tabela1}.

\section{Summary and discussion}\label{diskusija}

In this paper we have studied the configurational properties of linear polymers in the vicinity of adsorbing
wall of fractal containers modeled by self-avoiding walks on 3d SG
family of fractals. Each member of the 3d SG fractal family
has a fractal impenetrable 2d adsorbing boundary (which is,
in fact, 2d SG fractal surface) and can be labeled by an
integer $b$ ($2\le b\le\infty$).
 In this model interactions between monomers and adsorbing wall are described by parameters $w={\mathrm{e}}^{-\varepsilon_w/k_BT}$, and
$t={\mathrm{e}}^{-\varepsilon_t/k_BT}$, where $\varepsilon_w$ is the energy of a monomer lying on the adsorbing surface, and $\varepsilon_t$ is the energy
of a monomer in the layer adjacent to the surface. Depending of the values of $w$ and $t$ the polymer chain can exists in one of three possible states: adsorbed, desorbed and surface attached~\cite{Bouchaud,z15}.

For the first three members ($b=2$, 3 and 4) of the 3d SG fractal family we have
performed the exact RG analysis to evaluate the surface critical exponents $\gamma_{11}$, $\gamma_1$,and $\gamma_s$ that govern the numbers of all possible polymer configurations  with  both, one, and no ends grafted on the adsorbing boundary,
respectively. To evaluate specific values of  $\gamma_1, \gamma_{11}$, and $\gamma_s$, one needs to know the values of the end-to-end distance critical exponent $\nu$ and eigenvalues $\lambda_B$ and $\lambda_S$ of matrices ${\mathcal M_S}$ and ${\mathcal M_\gamma}$ defined by (\ref{c1d6}) and (\ref{lambdab}) respectively.  The exact values for $\nu$ and $\lambda_B$ have been already found \cite{z15,Dhar78,DharVannimenus,Knezevic87,z17}, for fractals with $b=2$, 3 and 4, and it has remained to calculate the values of  $\lambda_S$, which  requires knowledge of the  elements of the matrix ${\mathcal M_{S}}$.  The required elements are polynomials, that can be formed by enumeration and classification of SAWs, described by
restricted partition functions $C_1$, $C_2$, $D_1$, $D_2$ and $D_3$ (see figure \ref{fig2}), and we have found that this is feasible for $2\le b\le4$.
The obtained specific results for studied surface critical exponents $\gamma_{11}$, $\gamma_1$, and $\gamma_s$ are given in  table~\ref{tabela1}.

\begin{figure}
\hskip5cm
\includegraphics[scale=1.1]{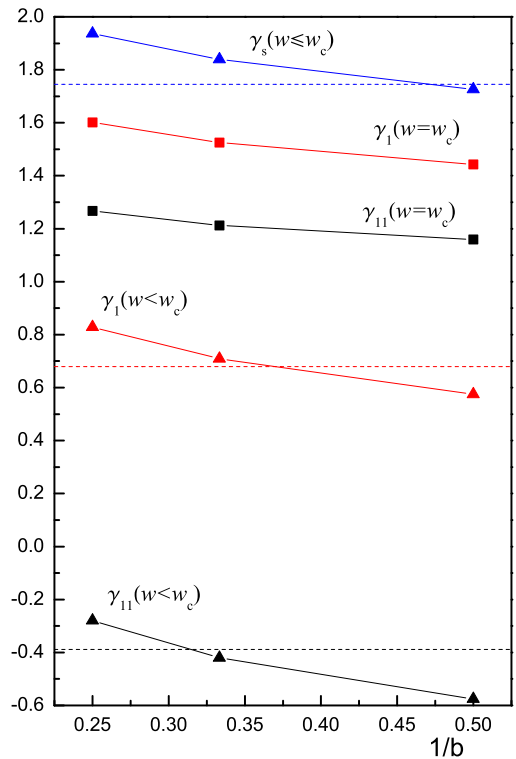}
\caption{ \label{fig:gamma}
 Values of the critical exponents $\gamma_{11}$, $\gamma_1$ and $\gamma_s$, for
desorbed ($w<w_c$) and surface attached polymer ($w=w_c$) as functions of $1/b$ ($b=2$, 3 and 4).  Full lines only serve as guides for the eyes.
The $d=3$ Euclidean values $\gamma_{11} \approx-0.389$, and $\gamma_{1} \approx 0.679$,  for desorbed polymer phase ($w<w_c$), that spring from two  findings \cite{diehl98,grassberger05}, are depicted by dashed horizontal lines, black for $\gamma_{11}$ and red for $\gamma_{1}$. The dashed horizontal blue line corresponds to the $d=3$ Euclidean value of $\gamma_s\approx 1.7449(2)$.}
\end{figure}
 For the sake of a better assessment of the global behavior
of the surface critical exponents as functions of the scaling
parameter $b$, we present our results  in figure~\ref{fig:gamma}. It can be seen that the following chain of inequalities is satisfied
\begin{equation}\label{inequalitychain}
\fl \gamma_{11}(w<w_c)<\gamma_1(w<w_c)<\gamma_{11}(w=w_c)<\gamma_1(w=w_c)<\gamma_s(w\leq w_c)\,.
\end{equation}
This is a quite plausible result  since  the generating function $G_{11}$
describes SAWs with a stronger constraint than those  described by
the generating function $G_{1}$,  which implies the inequality $\gamma_{11}<\gamma_1$, for both desorbed and surface attached chain.
Also, we see that for each $b$ the inequalities $\gamma_{11}(w<w_c)<\gamma_{11}(w=w_c)$, and $\gamma_{1}(w<w_c)<\gamma_{1}(w=w_c)$ are valid, implying that the number of SAW configurations with one  or two ends anchored to the wall are larger in the surface attached regime then in the desorbed one. Here we note that the  inequality $\gamma_{11}<\gamma_1$, for both desorbed and surface attached chain, were established for SAWs grafted on one edge of fractals belonging to  2d SG fractal family, whereas for $b>3$  it was obtained that $\gamma_{1}(w<w_c)>\gamma_{11}(w=w_c)$ \cite{EKMZ}.
Furthermore, for $w<w_c$, we  see that $\gamma_{11}$ and $\gamma_{1}$  are monotonically increasing functions of $b$, and can be compared with   their   Euclidean counterparts.  In desorbed phase  the exponent $\gamma_{11}$  for $b=4$ is larger than the corresponding three--dimensional Euclidean value $\gamma_{11}^{3d}\approx-0.389$  that springs from the field-theory approach result $\gamma_{11}^{3d}=-0.388$ \cite{diehl98} and Monte Carlo simulations finding $\gamma_{11}^{3d}=-0.390(2)$ \cite{grassberger05} (black dashed line in figure~\ref{fig:gamma}),
whereas $\gamma_{1}$
surpasses the Euclidean value $\gamma_{1}^{3d}=0.679$ (evaluated from $\gamma_{1}^{3d}=0.680$ \cite{diehl98} and $\gamma_{1}^{3d}=0.6786(12)$ \cite{grassberger05}), for $b>2$.
In the case of the surface attached extended chain ($w=w_c$),
both critical exponents, $\gamma_1$ and $\gamma_{11}$
are also monotonically increasing functions of $b$, being always larger then
the corresponding Euclidean values for $\gamma_{11}^{3d}\approx0.686$ (obtained as an average of $\gamma_{11}^{3d}=0.666$ \cite{diehl98}, and $\gamma_{11}^{3d}=0.707(5)$  \cite{grassberger05}) and $\gamma_{1}^{3d}\approx1.216$ (average of $\gamma_{1}^{3d}=1.207$ \cite{diehl98} and $\gamma_{1}^{3d}=1.226(2)$ \cite{grassberger05}). The behavior of the critical exponent $\gamma_s$
(whose values are the same in both the desorbed chain phase and the surface attached chain region), is similar to the behavior of $\gamma_1$ and
$\gamma_{11}$, that is, $\gamma_s$ is monotonically
increasing function of $b$, which  for $b>2$ surpasses its
Euclidean counterpart  $\gamma_s^{3d}\approx1.7449(2)$ (calculated from scaling relation $\gamma_s=\gamma+\nu$ \cite{Debell}, where we have put  the last estimates for three dimensional Euclidean values of  $\gamma=1.1573(2)$ \cite{Hsu} and $\nu_E= 0.587597(7)$~\cite{clisby10}).

Finally, we have tested the scaling relation
\begin{equation}\label{barber}
\gamma_s=2\gamma_1-\gamma_{11}=\gamma+\nu(d_f-d_s)\,,
\end{equation}
which was proposed in  \cite{Bubanja} as a modification of the scaling relation $2\gamma_1-\gamma_{11}=\gamma+\nu$, obtained for Euclidean containers~\cite{Barber,Debell}. Here $d_f$ is the fractal dimension of the solvent and $d_s$ is the fractal dimension of the attracting surface.
Putting the data  from  table~\ref{tabela1} into (\ref{barber}),
 one can see that this scaling relation  is exactly satisfied for both desorbed and surface attached polymers.

At the end, one may pose the question about the possible behavior of the surface critical exponents $\gamma_1, \gamma_{11}$, and $\gamma_s$ for larger $b$
(when the SG fractal dimension approaches the Euclidean
value 3).  To answer this question one have to apply some other method, such as  Monte Carlo renormalization group method which appeared to be very precise and quite efficient method for critical exponents calculation on finitely ramified    fractals,  which will be the matter of an independent study.

\ack{This paper has been done as a part of the
work within the project No. 171015 funded by the Serbian Ministry of Education and Science.}

\appendix

\section{\label{appendixa} RG equations for the surface one-leg parameters for $b=2$ 3d SG fractal\label{dodatak1}}

In this Appendix we give the exact RG equations (\ref{c1d6}) for the surface one-leg
parameters $C_1$, $C_2$, $D_1$, $D_2$ and $D_3$ in the case of the  $b=2$ 3d SG
fractal. We have found that these equations have the following
form:
\begin{eqnarray}
\fl C_1'=(3 A + 6 A_2^2 B)C_1+(6 A A_2 + 6 A A_1 A_2)C_2+ 6 A_2^2 B\, D_1+ (6 A A_1^2 + 12 A_2^2 B)D_2\,
,\nonumber \\
\fl C'_2=(2 A A_2 + 2 A A_1 A_2)C_1+ (1 + 2 A_1 + 2 A_1^2 + 2 A A_2^2 + 4 A A_2 B_1)C_2+\nonumber \\
  (2 A A_2 + 4 A A_1 A_2 + 4 A A_1 B_1)D_1+(2 A A_2 + 4 A A_1 A_2 + 8 A A_1 B_1)D_2+\nonumber \\
  (2 A_1^2 + 2 A A_2^2)D_3   \, ,\nonumber \\
\fl  D_1'=(A A_1^2 + A_2^2 B)C_1+ 2 A A_1 B_1\, C_2+
  (2 A A_1 + 3 A A_1^2 + 2 A_2^2 B + 8 A_2 B B_1 + 10 B B_1^2)D_1+\nonumber \\
  (2 A_2^2 B + 4 A_2 B B_1 + 12 B B_1^2)D_2+ (2 A A_1 A_2 + 4 A A_1 B_1)D_3\,,\nonumber \\
 \fl   D_2'=A_2^2 B\, C_1+
  (A A_2 + 2 A A_1 A_2 + 2 A A_1 B_1)C_2+ (2 A_2^2 B + 4 A_2 B B_1 + 6 B B_1^2)D_1+\nonumber \\
  (2 A A_1 + 3 A A_1^2 + 4 A_2^2 B + 12 A_2 B B_1 + 16 B B_1^2)D_2
  +(2 A A_1 A_2 + 6 A A_1 B_1)D_3\,,\nonumber \\
\fl   D_3'=(A_1^2 + A A_2^2)C_2+(2 A A_1 A_2 + 6 A A_1 B_1)D_1+
  (4 A A_1 A_2 + 10 A A_1 B_1)D_2+\nonumber \\ (3 A_1^2 + 22 A B_1^2)D_3 \,. \nonumber
 \end{eqnarray}


\end{document}